\documentclass[openany, 12pt]{article}
\usepackage[utf8]{inputenc}
\usepackage[a4paper,top=3.cm,bottom=3.cm,left=3.cm,right=3.cm,marginparwidth=1.75cm, headheight=15pt]{geometry}
\usepackage{setspace}
\usepackage{amsmath}
\usepackage{amsfonts}
\usepackage{mathrsfs}
\usepackage{cases}
\usepackage{graphicx}
\usepackage{caption}
\usepackage{subcaption}
\usepackage{appendix}
\usepackage{hyperref}
\usepackage{longtable}
\usepackage{fancyhdr}
\usepackage{appendix}
\usepackage{breakcites}
\usepackage{authblk}

\usepackage{abstract}

\hypersetup{
    colorlinks=true,
    linkcolor=blue,
    citecolor=blue,
    urlcolor=black
    }

\title{Understanding and nowcasting the illicit drug distribution in England: a data-centric approach to the County Lines Model}
\author[1, 2, $\star$]{Leonardo Castro-Gonzalez}
\affil[1]{University of Birmingham, Edgbaston, UK}
\affil[2]{The Alan Turing Institute, London, UK}
\affil[$\star$]{l.m.castrogonzalez@bham.ac.uk}
\date{}

\begin{document}

\maketitle

\begin{abstract}
The County Lines Model (CLM) is a relatively new illicit drugs distribution method found in Great Britain. The CLM has brought modern slavery and public health issues, while challenging the law-enforcement capacity to act, as coordination between different local police forces is necessary.
Our objective is to understand the territorial logic behind the line operators when establishing a connection between two places. 
We use three different spatial models (gravity, radiation and retail models), as each one of them understands flow from place $i$ to $j$ in a different way. Using public data from the Metropolitan Police of London, we train and cross-validate the models to understand which of the different physical and socio-demographic variables are considered when establishing a connection. We analyse hospital admissions by drugs, disposable household income, police presence and knife crime events, in addition to the population of a particular place and the distance and travel times between two different. 
Our results show that knife crime events and hospital admissions by misuse of drugs are the most important variables. 
We also find that London operators distribute to the territory known as the ``South’’ of England, as negligible presence of them is observed outside of it
\end{abstract}

\section{Introduction}
\label{sec:introduction}

During the last decade, a new illicit drugs distribution model has been developed in the UK. The model was baptised as the "County Lines Model" (CLM) by the UK government \cite{NCA2018} given its use of phone lines that are established between different counties. The problem has become increasingly worrying each year, becoming a top priority for security agencies given the limited ability to stop them, and the modern slavery and public health problems that the CLM brings to local communities \cite{NCA2018, Black2020_evidence, Andell2018, Robinson2019, Stone2018, CAMBER2020, Moyle2019}. 

The \textit{modus operandi} can be described in the following way: a central hub is settled in big English cities like London, Birmingham, Manchester and Liverpool, from where drugs are sold and distributed \cite{Coombes2018}. From these hubs, \textit{lines} are settled to other parts of the country where a local market is established. So-called \textit{settlers}, find a local accommodation (normally a flat belonging to consumers) in the destination market from which drugs can be distributed. Local \textit{runners} are then hired to distribute the illicit merchandise to the consumers. Runners tend to be young people with knowledge of the local market whose tasks are to deliver merchandise and attract new clients. The distribution model increases the efficiency with respect to ``the traditional model'' \cite{Coombes2018} where the ``highstreet'' illicit drug seller buys merchandise to a bigger distributor, to then sell it on the street. The improvement of the CLM is to merge both tiers (local and bigger seller) uniting both channels of distribution (hub-settler and runner-consumer). 

Local consumers are given a phone number where they can place an order. The call is normally picked up in the central hub, from where they make the arrangements to distribute it to the consumer via the settler and the runner. The settler travels back and forth from the central hub and the local market bringing merchandise, while the runner distributes to the final consumer. 

According to the National County Lines Coordination Centre \cite{NCLCC2021}, 3 cities account for more than 80\% of the detected County lines in Great Britain in 2019 and 2020. These are, in respective importance, London, Birmingham and Liverpool. Public data is scarce, only having detailed records for London for those two years.\\

The implications of the proliferation of the CLM in the UK are multiple. Three are particularly highlighted in the literature \cite{Andell2018, Robinson2019, Black2020_summary}: (1) the rising of new illicit drug markets in small coastal towns and rural areas of England where illicit drugs problem were not found before.
(2) Also, the involvement of young vulnerable people in the distribution scheme is of worrisome for the UK Gov. This population is the most prone to be caught by law-enforcement bodies, while being involved in a modern slavery scheme making them hard to leave the CLM once they are involved.
(3) Finally, a limited ability from the different police forces in England to dismantle any complete distribution channel between one place and another. Cooperation between different law-enforcement bodies is necessary, as every link in the distribution engine can work autonomously, making it hard for bodies to dismantle the whole distribution operation. 

The fact that county line operators are found in small villages and coastal towns, far from local capitals and larger cities has risen different hypotheses about the logic behind establishing a \textit{line}. Indeed, the size of the population of the target places seems not to be a primordial element, as large population centres (London, Manchester, etc.) do not attract a big number of lines according to public data shown in the strategic report from the National County Lines Council \cite{NCLCC2021}. According to the same report, the logic behind the gangs operating county lines is a supply-demand balance. \\ 

The main objective of this work is to understand the \textit{territorial logic} that county line operators follow to establish different distribution routes. To do so, we have to answer the question if the ``traditional distribution model'' has been broken as literature suggests \cite{Stone2018}. If so, which are the new social, demographic and economic elements that are now taken into account to establish a new route? To answer both questions would help to obtain useful information for the Metropolitan Police to understand and tackle the county lines problem.

In order to answer the questions stated above, we test three different spatial interactions models to compute flows from one place $i$ to a second place $j$. We understand each of these models as different ways to understand the flow of persons/merchandise. Thus, by testing and comparing them we can extract information about which mechanisms could county lines operators follow. The models we use are the Gravity Model \cite{Anderson2010}, the Radiation Model \cite{Simini2012} and the Retail Model \cite{Wilson2008}, taking a similar approach as in \cite{Piovani2018} while extending it and adapting it to accomplish our objectives.

We use the classic gravity model as our benchmark, as it understands the flow from one place to another as proportional to the respective populations and inversely proportional to the distance between both places. Thus, we use it as a proxy of the traditional idea stating that more population would translate into more demand for illicit drugs. In that sense, a more populated city like Birmingham or Manchester would be more attractive for county lines operators than other rural places at the same respective distances from a given origin.

Radiation model understands flows as a process of sorting the available opportunities between $i$ and $j$. To arrive to place $j$, the studied element (person/merchandise) should not be captured by the opportunities found in the way to it. As an extension of our benchmark, in this case we are interested in testing the \textit{distribution of population} in England. That is, for example, not only taking into account the population of Birmingham and London, but also the population found in between. 

Finally, the retail model understands flow as a balance between the opportunities and the costs of going from one place to another, compared with all the other competing places in the given space.
This latter model allows to test other kind of dynamics involving different benefits and costs while considering competition too. The different benefits and costs can be of physical nature (time, distance), but also social or demographic. We explore five different independent variables we expect to have some leverage for operators. These are knife crime events, number of police officers, gross disposable household income, hospital admissions by misuse and poisoning by drugs as possible costs. 

The hospital admissions are taken as proxies for illicit drugs consumption, as no other data is available. In that sense, we are exploring correlations between other social elements that might be of a higher importance for county lines operators to establish local markets. 
Knife crime events are another high-priority incidents for UK Government \cite{Commons2019}, which are reported to be related to gang rivalry. We are interested to see if the presence of this kind of event could be an element taken into account for operators as a disincentive for establishing a local market. 
In the same way, we are expecting police workforce to be a disincentive for gangs. Finally, we take the gross disposable household income as a measure of richness, as average income does not take into account regional disparities in rent prices, money transfers from the government and local taxes.
We train and test the three models with public data from the Metropolitan Police of London \cite{Met2019} accounting the detected lines in other police force territories in Great Britain from London in 2019 and 2020. \\


In the following, we present a small literature review in Section \ref{sec:literature_review}. The different models and data tested in Section \ref{sec:methods}. Results are presented in Section \ref{sec:results}, to then discuss and conclude in Section \ref{sec:discussion}. We also present two Appendices \ref{sec:appendix_a} and \ref{sec:supplementary_matieral}. The former is a table to help the reader with the models tested, while the latter details the different sources and formats of the data used in this work.
To the authors' knowledge, this is the first published work that studies the County Lines Model from a quantitative approach.

\subsection{Related work}
\label{sec:literature_review}

 In the case of the County Lines Model, only qualitative and official literature has been published. The Official literature includes documents and reports from different police agencies and the UK government. In particular, the NCA has published each year a statement regarding the views of the organisation about the County Lines Model \cite{NCA2018}. The document presents the findings from the NCA to understand the model and the different consequence it has had in the population.
 
 In 2019, the UK Government's Home Office commissioned an up-to-date report to be done around the illicit drugs problem in the UK. The report was published in early 2020 \cite{Black2020_evidence, Black2020_summary} and reveals how the County Lines Model has evolved over the last decade. It also reports how the consumption of illicit drugs has changed in the population, stating that the UK faces an important challenge, as there currently are two peaks of consumers: one in their 20's and another in their 60's. Each one of those is of increasing worrisome, as the first one is the future workforce of the UK and the second represents an increasing pressure in the public services. 

Two different police organisations have publicly published information about the County Lines Model information they have. These are the Metropolitan Police of London \cite{Met2019, Met2020} and the West Midlands Police (Birmingham and metropolitan area) \cite{WestMidlands2020}. Only the Metropolitan Police has published quantitative data about their detection of lines in other police territories.

In January 2018, a debate was held in the House of Commons (UK's lower parliamentary chamber) to discuss the exploitation and harms done by the County Lines Model in London \cite{Commons2018}. Different Members of the Parliament asked what has been done until that point to tackle the CLM problems in London, particularly gang activity and exploitation.

Outside official documentation, academic literature about County Lines has mostly dedicated to report the child exploitation in different locations of England \cite{Andell2018, Robinson2019, Stone2018} and Scotland \cite{Gavin2018}. In all of them we find a description of the model. 
An anthropological study can be found in \cite{Ross2017}, where the authors interview different consumers and victims of the CLM in South England. \\

The present research is also found in the current context of need for better information for law-enforcement bodies in the UK, as there is an ongoing discussion about how Brexit and the COVID-19 pandemic will have a major effect on public spending, particularly in law enforcement bodies and the National Health Services (NHS, the public health body in the UK) \cite{Roman-Urrestarazuk4003}. 
In particular, reports state historical maximum numbers of drug-related deaths \textit{per capita}, as a new generation of young consumers enters the market and an older generation  requires more health care services \cite{Black2020_evidence}.
Also, it has been discussed how Brexit would make more difficult for the United Kingdom to access and profit from European funding and infrastructure (like the European Monitoring Centre for Drugs and Drugs Addiction, EMCDDA) for better intel and tackling strategies for a better public health and general quality of life for its citizens \cite{Coombes2018}. 
\section{Methods}
\label{sec:methods}

We cannot speak of a flow of persons, but rather a number of detected \textit{lines} (connections) established from a place $i$ to another place $j$. In that sense, the data point is a natural number, $T_{ij}^\mathrm{data}$, representing the detected connections. 

The spatial resolution we work with is at police force territory, which in Great Britain account for 39 in England, 5 in Wales and 1 in Scotland. In our case, we work with the 39 territories in England only to train our models. We only train for England as not all features used in the models are available for the whole of Great Britain. We merge both territories in Greater London (Metropolitan Police + City of London Police) to work with London as a unique space.

\subsection{Retail model}
\label{ssec:retail}

The Retail model was first presented in \cite{Wilson2006} as an entropy-maximising model for the function $T_{ij}^\mathrm{retail}$ with three different conditions: (a) an outflow condition $\sum_jT_{ij}^\mathrm{retail}=T_i$. (b) a Boltzmann-inspired energy conservation condition with respect to the travel time $c_{ij}$ from $i$ to $j$, $\sum_{i,j}T_{ij}^\mathrm{retail}c_{ij}=C$, and (c) a similar conservation condition with respect to the total \textit{benefit} found in the space, $\sum_{ij}T_{ij}^\mathrm{retail}\log w_j=B$, where $w_j$ is the benefit of place $j$ to attract people. 

Using the maximum entropy principle with the three constraints described above, we obtain the resulting function for $T_{ij}^\mathrm{retail}$:
\begin{equation}
    \label{eq:BLV_original}
    T_{ij}^\mathrm{retail}=\frac{T_i\exp\{\alpha\log w_j-\beta c_{ij}\}}{\sum_k\exp\{\alpha\log w_k-\beta c_{ik}\}}.
\end{equation}
Where $\alpha$ and $\beta$ are two free parameters coming from the maximum entropy derivation. Notice how the exponent in the numerator represents the balance from the benefits at $j$ and the cost to get to $j$ from $i$, given by $\alpha\log w_j-\beta c_{ij}$. This latter balance competes with the other balances of going to the places $k$ via the denominator of Eq. (\ref{eq:BLV_original}).

The retail system has been studied for different spatial dynamics in the past \cite{Piovani2018, Davies2013}, allowing to include different types of data as benefit $w_j$. In this case, as we are interested in knowing if different social variables (police workforce, knife crime events, hospital admissions by misuse of or poisoning by drugs and drug-related deaths) might be relevant benefits or costs for the county lines operator, we thus replace condition (c) mentioned above by 5 analogous restrictions, one per variable, and use the different $w_j^{(n)}$ as the social/demographic variables. All of them (gross disposable household income, police workforce, knife crime events and hospital admissions) are normalised by the population of the police territory so they become per 100 000 inhabitants. We thus obtain as final solution
\begin{equation}
    \label{eq:BLV_modified}
    T_{ij}^\mathrm{retail}=\frac{T_i\exp\{\sum_n\alpha_n\log w_j^{(n)}-\beta c_{ij}\}}{\sum_k\exp\{\sum_n\alpha_n\log w_k^{(n)}-\beta c_{ik}\}}.
\end{equation}
By exploring the magnitude and sign of the different $\alpha_n$, we can then have an insight about the elements that correlate to the detected lines from the Metropolitan Police, and if the variable is perceived as a benefit ($\alpha_n>0$) or a cost ($\alpha_n<0$).

\subsection{Gravity model}
\label{ssec:gravity}

The gravity model computes flows from $i$ to $j$ as proportional to the product of populations of $i$ and $j$, and inversely proportional to the distance between them. The model has different expressions and different limitations \cite{Simini2012, Noulas2012}. We take as basis for this work the following form \cite{Anderson2010}
\begin{equation}
    \label{eq:gravity_original}
    T_{ij}^\mathrm{gravity}=G\frac{m_i^a m_j^b}{d_{ij}^c}.
\end{equation}
We impose the outflow restriction $\sum_j T_{ij}^\mathrm{gravity}=T_i$, which makes Eq. (\ref{eq:gravity_original}) become
\begin{equation}
    \label{eq:gravity_modified}
    T_{ij}^\mathrm{gravity}=T_i\bigg(\sum_{k\neq i}\frac{m_k^b}{d_{ik}^c}\bigg)^{-1}\frac{m_j^b}{d_{ij}^c}.
\end{equation}

\subsection{Radiation model}
\label{ssec:radiation}

The idea behind the radiation model originally comes from a particle transmission and absorption model in physics, where a particle is supposed to be emitted from place $i$ and arriving to place $j$ by sorting all \textit{opportunities} in the way, i.e. not being absorbed in the way from one place to another. This idea has been applied for flow of persons in a given space, first used as a commuter model for job seeking in the US \cite{Simini2012}, to then being applied into different examples where commuters are modelled \cite{Piovani2017, Masucci2013}. The original formulation of the radiation model is
\begin{equation}
    \label{eq:radiation_original}
    T^\mathrm{rad}_{ij} = T_i\frac{p_ip_j}{(p_j+p_{ij})(p_i+p_j+p_{ij})},
\end{equation}
where $p_i$ and $p_j$ are the populations of $i$ and $j$, $p_{ij}$ is the sum of populations between both places and $T_i$ is given by the outflow constraint $T_i = \sum_{j\neq i}T_{ij}^\mathrm{rad}$. In this particular project we work with a modified version from \cite{Yang2014}:
\begin{equation}
    \label{eq:radiation_modified}
    T_{ij}^\mathrm{rad}=T_i\frac{P(1|n_i, n_j, n_{ij})}{\sum_kP(1|n_i, n_k, n_{ij})},
\end{equation}
where $n_i$, $n_j$ and $n_{ij}$ are the opportunities in $i$, $j$, and between both places respectively. In this case, we simply suppose that $n_i=\rho p_i$, with $P(1|n_i, n_j, n_{ij})$ as the probability of the ``particle'' being absorbed in way from $i$ to $j$ given the opportunities $n_i$, $n_j$ and $n_{ij}$.
\begin{equation}
    \label{eq:radiation_probs}
    P(1|n_i, n_j, n_{ij})=\frac{[(n_i+n_j+n_{ij})^r-(n_i+n_{ij})^r](n_i^r+1)}{[(n_i+n_{ij})^r+1][(n_i+n_j+n_{ij})^r+1]}.
\end{equation}

\subsection{Model selection process}
\label{ssec:loss}

The three models presented above represent different spatial interactions, interpreted in this context as different decision processes from the county lines operators to establish a connection between place $i$ and $j$.
To compare the different models and selecting the most appropriate one for our available data, we proceed using two different measures found in the literature: the S\o{}rensen-Dice index $S$ \cite{Piovani2017}, and the Bayesian information criterion (BIC) which is based in the maximum likelihood principle \cite{Altmann2020}. \\

The S\o{}rensen-Dice $S$ index measures the similarity between two different samples. Given a modelled number of detected lines $T_{ij}^\mathrm{model}$ after any of the models described above, and the observed data $T_{ij}^\mathrm{data}$, we use the same formulation as in \cite{Piovani2017}
\begin{equation}
    \label{eq:sorensen}
    S = \frac{2\sum_{i,j}\min(T_{ij}^\mathrm{data}, T_{ij}^\mathrm{model})}{\sum_{i,j}T_{{ij}}^\mathrm{data}+ \sum_{i,j}T_{ij}^\mathrm{model}}.
\end{equation}
We perform a 2-fold cross-validation, splitting our database for 2019 and 2020. Thus, training with 2019 (2020) data to then validate with 2020 (2019) data. 
The main argument around why we perform a 2-fold cross-validation, and not an $n$-fold one with a higher $n$ is that, in order to comply with an accurate comparison between the different models, the cross-validation must be performed in the same folds for each of the models. By including the radiation model in Eq. (\ref{eq:radiation_modified}) which works in slices of land rather than individual points, we would then have to correctly choose our different folds, so no information is lost when slicing. However, given the topology of England and the way the variable $n_{ij}$ is constructed for Eq. (\ref{eq:radiation_modified}), we could only slice England in two different pieces, which by themselves are not well balanced (the south-east of England, and the rest of the country).

As an extra criterion to model selection, we also compute the BIC to the whole modelled sample by each of the models. BIC computes the log-likelihood and corrects with the size of the sample $M$ for each model. In that sense, 
\begin{equation}
    \label{eq:aic}
    BIC = 2\log M-2\log\hat{L}.
\end{equation}
$M$ is the size of the sample and $\log\hat{L}$ represents the maximum value obtained for the log-likelihood when training the model. The log-likelihood is computed with the parameters that minimise the loss functions used to calibrate the model. 

As discussed before, given the nature of the detected lines by the Metropolitan Police, we are interested in testing two different loss functions: the usual mean-square loss function derived from a Gaussian likelihood, shown in Eq. (\ref{eq:loss_gauss}), and a loss function derived from a Poissonian likelihood, shown in Eq. (\ref{eq:loss_poisson}). The chose of the Poissonian likelihood is given by the distribution of lines detected for both years, while the mean-square loss function is chosen to be a benchmark with respect to Eq. (\ref{eq:loss_poisson}).
\begin{numcases}{}
    \label{eq:loss_gauss}
    $$\mathcal{L}_\mathcal{G}\Big(\{T_{Lj}^\mathrm{model}(\hat{\theta})\}_j\ | \ \hat{\theta}\Big)=\frac{1}{2N}\sum_j\Big(T_{Lj}^\mathrm{data}-T_{Lj}^\mathrm{model}\Big)^2,$$ \\
    $$\mathcal{L}_\mathcal{P}\Big(\{T_{Lj}^\mathrm{model}(\hat{\theta})\}_j \ | \ \hat{\theta}\Big)=\frac{1}{N}\sum_jT_{Lj}^\mathrm{model}-T_{Lj}^\mathrm{data}\log \  T_{Lj}^\mathrm{model},$$ \label{eq:loss_poisson}
\end{numcases}
where $\hat{\theta}$ is the vector of free parameters for each model. To each of both loss functions we are also adding an L2 regularisation term $\lambda||\hat{\theta}||^2$, with $\lambda=1$. The subscript $L$ in $T_{Lj}$ represents London, thus showing the observation/model for London to any other police territory $j$.\\


\subsection{\label{sec:pipeline}Pipeline}
The analysis pipeline is as follows: we perform a 2-fold cross-validation on each of the three types of models (gravity, radiation and retail). In total, we are training 1 gravity model, 1 radiation model and 32 retail models. The 32 retail models are a result of adding an offset to the 5 different free parameters $\{\alpha_n\}$ included in the Retail model of Eq. (\ref{eq:BLV_modified}). Thus, the total number of models is $\sum_{i=0}^{5}\binom{5}{i}=32$. For all the 32 models we still take into account the $\beta$ parameter which accounts for the travel times cost. A more detailed list of the models trained can be found in Appendix \ref{sec:appendix_a}.
The models are trained using two different cost functions described in Section \ref{ssec:loss}, and evaluated using the S\o{}rensen-Dice index \cite{Piovani2017} and the Bayesian Information Criterion (BIC) \cite{Altmann2020}.


\subsection{Data}
\label{ssec:data}

In this subsection we describe the different data that is implemented in the different tested models. In Appendix \ref{sec:supplementary_matieral} we offer a more detailed description of the complete database used. The three models (gravity, radial and retail) have as one of the inputs the population of the police territories (directly or indirectly). These are public data from the Office of National Statistics (ONS), and by the time of submission, the last published update is of 2019\footnote{In general, all data obtained from the digital platforms of the British Government (gov.uk) is used under the Open Digital Licence.}. 

The Gravity and the Retail model respectively use the distance and the travel time from one place to another. Given that the used resolution is at police territory level, we are using the distance/travel time from the most populous place in territory $i$ to the most populous place in $j$. Data is drawn using the Google Maps\copyright API. \\

The exponent of Eq. (\ref{eq:BLV_modified}) allows to compute a balance between the different benefits and costs of going from $i$ to $j$. The training and comparison process taken in this work allows to know if a given variable is a cost or a benefit, thus allowing to test between different variables.

An important feature to test is the amount of potential costumers for the county lines operators. This accounts to current and potential consumers. We use two different measures as proxy to this consumption: finished hospital admissions \cite{NHS2019} by misuse of drugs and finished admissions by poisoning of drugs. 
Hospital admissions are normalised by population and by daytime hospital beds \textit{per capita}.

Another feature we test is the police workforce in each territory. We use the number of average Full-time police officer over the British Fiscal year (May-April) which can be obtained from \cite{HomeOfficeWorkforce2019}. 

To account for the disparities of richness in the different parts of England, we use the gross dispensable household income (gdhi). In comparison with the household income, the gdhi takes into account the amount of money that households have after local and national income taxes and benefits from the government. Data was obtained from the ONS \cite{GDHI2021}.

Finally, we are interested in testing the knife crime events \textit{per capita} in each of the police territories. Knife crime events have been an increasingly worrying matter for the British Government, with numbers increasing 78\% in England from 2014 to 2020 \cite{Commons2019}.


\section{Results}
\label{sec:results}

\subsection{Model selection}
\begin{figure}[ht]
    \centering
    \begin{subfigure}{0.49\textwidth}
        \centering
        \includegraphics[width=\textwidth]{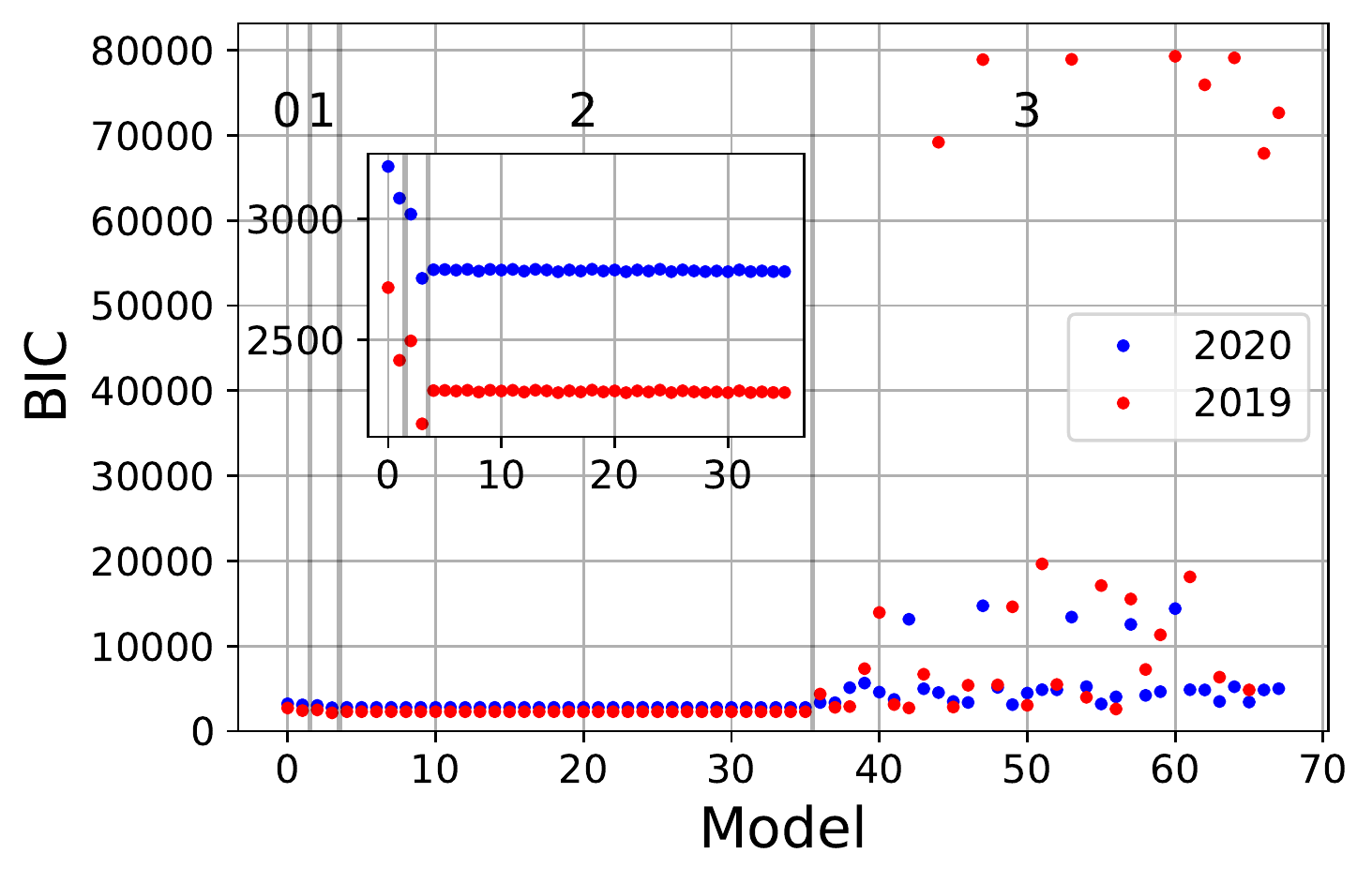}
        \caption{\label{fig:BIC}}
    \end{subfigure}
    \begin{subfigure}{0.49\textwidth}
        \centering
        \includegraphics[width=\textwidth]{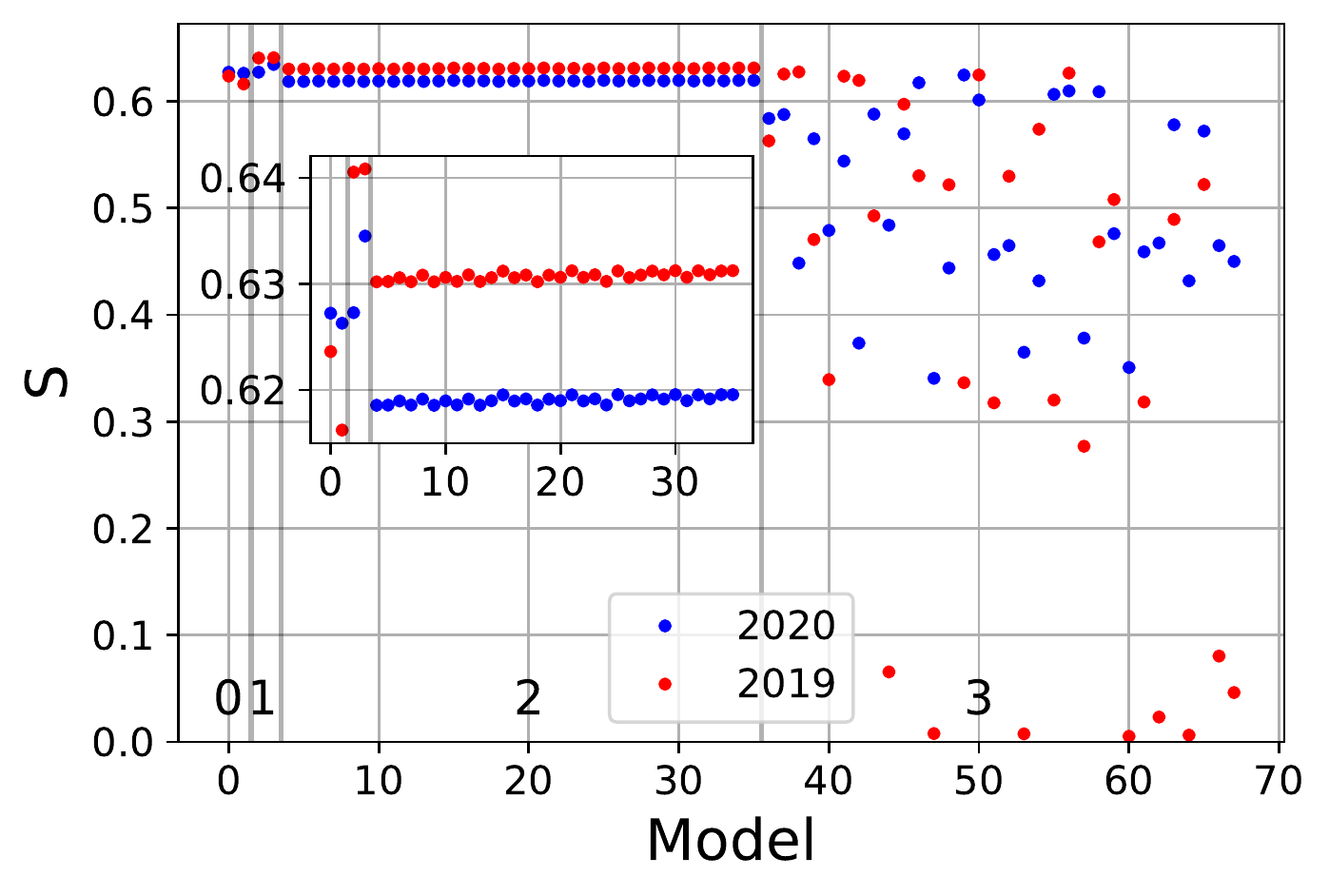}
        \caption{\label{fig:Sorensen}}
    \end{subfigure}
    \caption{\label{fig:indices}Results of BIC and the S\o{}rensen-Dice index for the 68 different models tested. Zone 0 corresponds to the Gravity model. Zone 1 corresponds to the Radiation model. Finally, zones 2 and 3 correspond to the retail model with the Poissonian loss function and with the MSE loss function respectively. The annotated year corresponds to the data in which the model was trained on. Details of each model can be found in Appendix A.}
\end{figure}

Results for the BIC and the S\o{}rensen-Dice index are found in Figure \ref{fig:BIC} and in Figure \ref{fig:Sorensen} respectively. 

When comparing the Retail model calibrated with a Gaussian loss function (zone 3 in Figure \ref{fig:indices}) with respect to the other models, we observe how the former performs worst in all of its forms for both the BIC and the $S$ index. We can thus proceed to discard these models.

\begin{figure}
    \centering
    \includegraphics[width=0.5\textwidth]{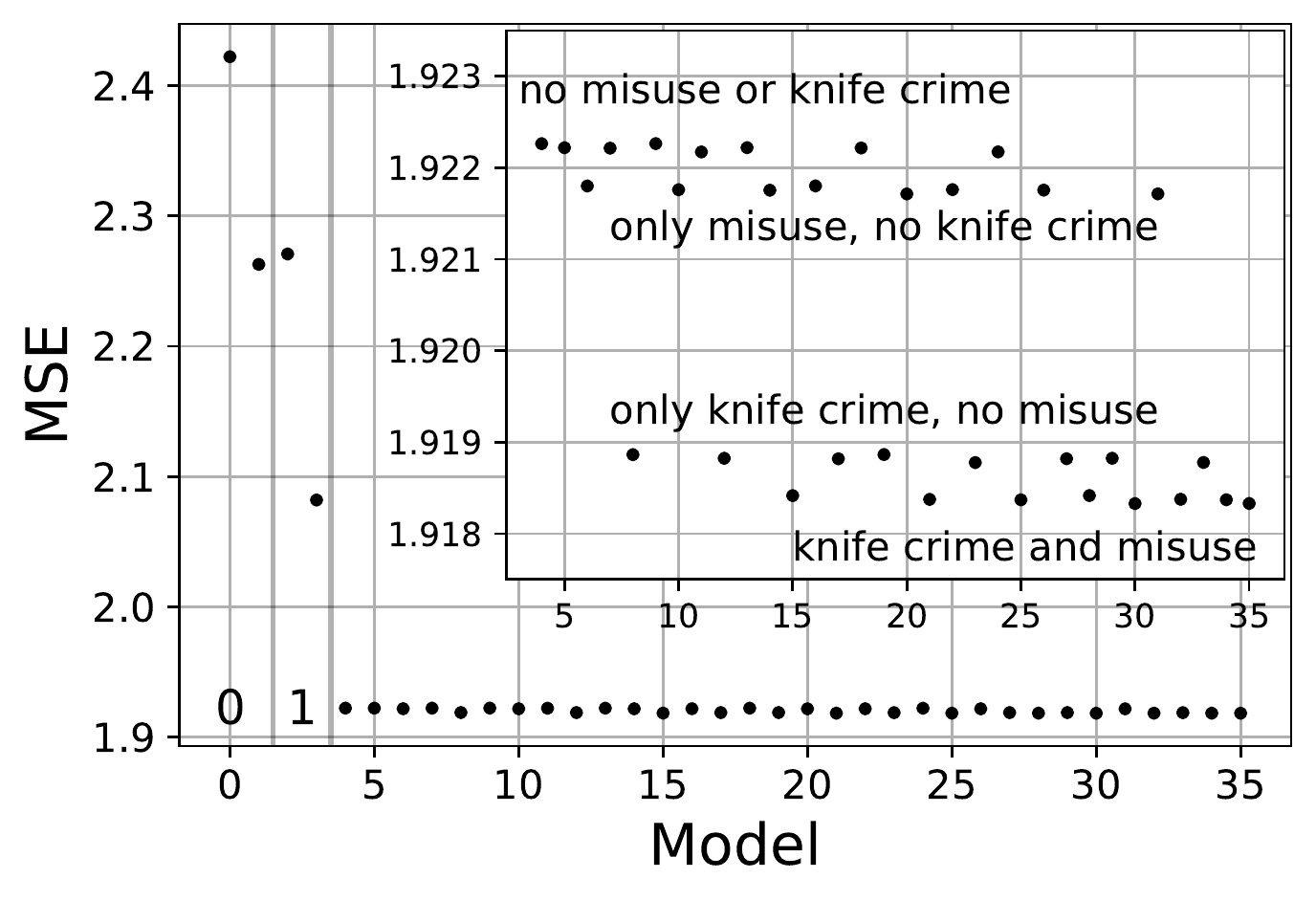}
    \caption{MSE costs when comparing the trained models with the Metropolitan Police data.}
    \label{fig:costs}
\end{figure}

With the models left, we perform a comparison by computing the MSE between the Metropolitan Police data for both years (2019 and 2020) and the predictions obtained from each model. The MSE is computed with the logarithms of the data points, so in this case $MSE = \frac{1}{N}\sum_{i,j}(\log T_{ij}^\mathrm{data}-\log T_{ij}^\mathrm{model})^2$ Results are seen in Figure \ref{fig:costs}. 

The best performing model is the Retail model trained with the 2019 data and the Poissonian loss function. However, as it can be seen in the inset plot in Figure \ref{fig:costs}, the results can be differentiated in 4 levels. When examining each one of thems, we find that the hospital admissions by poisoning of drugs, the disposable income and the police presence variables do not have significant effect on the performance of the model. This can be seen in the upper level, as those combinations not containing the knife crimes and hospital admissions by misuse of drugs variables are those present there (all the different models are in Appendix \ref{sec:appendix_a}). The fact that the combination without any of the social variables and only the travel times is there allow us to interpret that any of the three mentioned variables before do not have any particular effect on the performance of the model. 
The hospital admissions by misuse of drugs seem to have an impact on the cost, although not as important as the knife crime variable. When combining both variables we obtain the most important effect on the MSE cost and the best performing models.

The Radiation model follows as best performing when trained with the 2019 data and Poisson loss function. Finally, we obtained the Gravity model trained in the same way. \\

In Table \ref{tab:results_calibration} we detail all the selected models. 
To keep the selected models as simple as possible, we filter out all the different Retail models and keep only those with the minimum number of variables. That is, one with both the misuse and the knife crime variables in addition to the travel times, one with only the knife crime variable and travel times, one with only the misuse variable and travel times, and finally one with only travel times. 

From the exponent in Eq. (\ref{eq:BLV_modified}), 
$\alpha_2$ corresponds to the hospital admissions by misuse of drugs, 
 and $\alpha_4$ to the knife crime events.
 All variables are normalised by population. 

We also select the best performing Radiation and Gravity models as we are interested in comparing them with respect to the Retail model. \\

As it can be seen from Table \ref{tab:results_calibration}, the five exponents $\alpha_n$ are negative, which is interpreted as all of the variables to represent a cost to county lines operators. This will be discussed in Section \ref{sec:discussion}.

\begin{table}[ht]
    \centering
    \caption{\label{tab:results_calibration}Results for the best three models calibrated.}
    \hspace*{-.5cm}
    \begin{tabular}[t]{c|c|c|c|l}
    
         Ranking & Model & Loss function & Training year & Parameters \\
         \hline
        1 & Retail & Poisson & 2019 & \shortstack[l]{$\alpha_2=-0.774\mathrm{e}{-2}$, $\alpha_4=-0.013$,\\$\beta=0.014$}  \\
        \hline
        2 & Retail & Poisson & 2019 & \shortstack[l]{$\alpha_4=-0.013$, $\beta=0.014$}  \\
        \hline
        3 & Retail & Poisson & 2019 & \shortstack[l]{$\alpha_2=-0.777\mathrm{e}{-2}$, $\beta=0.014$} \\
        \hline
        4 & Retail & Poisson & 2019 & \shortstack[l]{$\beta=0.014$}\\
        \hline
        5 & Radiation & Poisson & 2019 & \shortstack[l]{$\rho=2.085$, $n=1.038$}\\
        \hline
        6 & Gravity & Poisson & 2019 & \shortstack[l]{$b=0.697$, $c=0.368$} \\ 
        \hline
    \end{tabular}
\end{table}

\subsection{Model analysis and geographic distribution}
\begin{figure}
    \centering
    \begin{subfigure}[b]{0.49\textwidth}
        \centering
        \includegraphics[width=\textwidth]{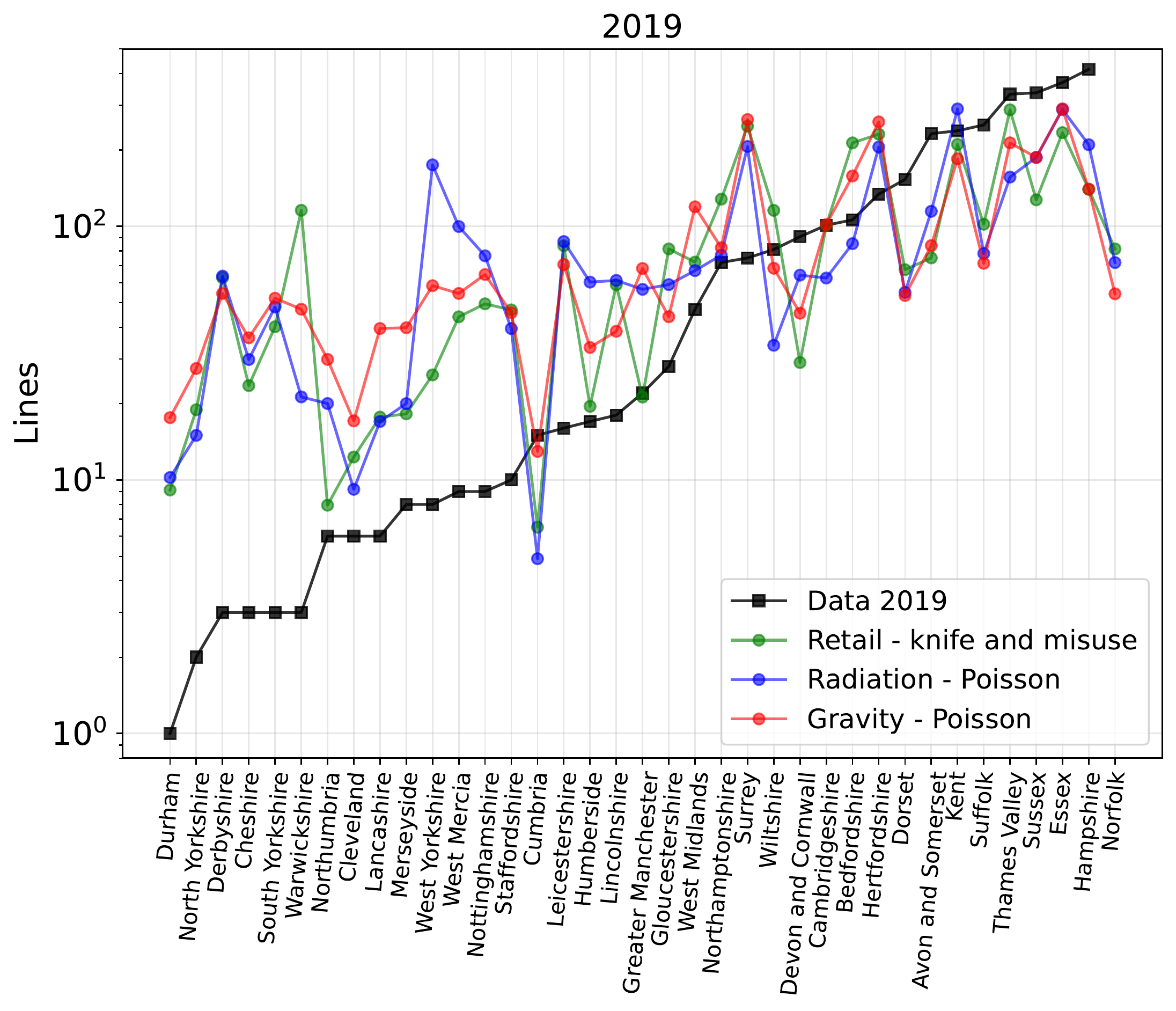}
        \caption{\label{fig:dispersion_19}}
    \end{subfigure}
    \begin{subfigure}[b]{0.49\textwidth}
        \centering
        \includegraphics[width=\textwidth]{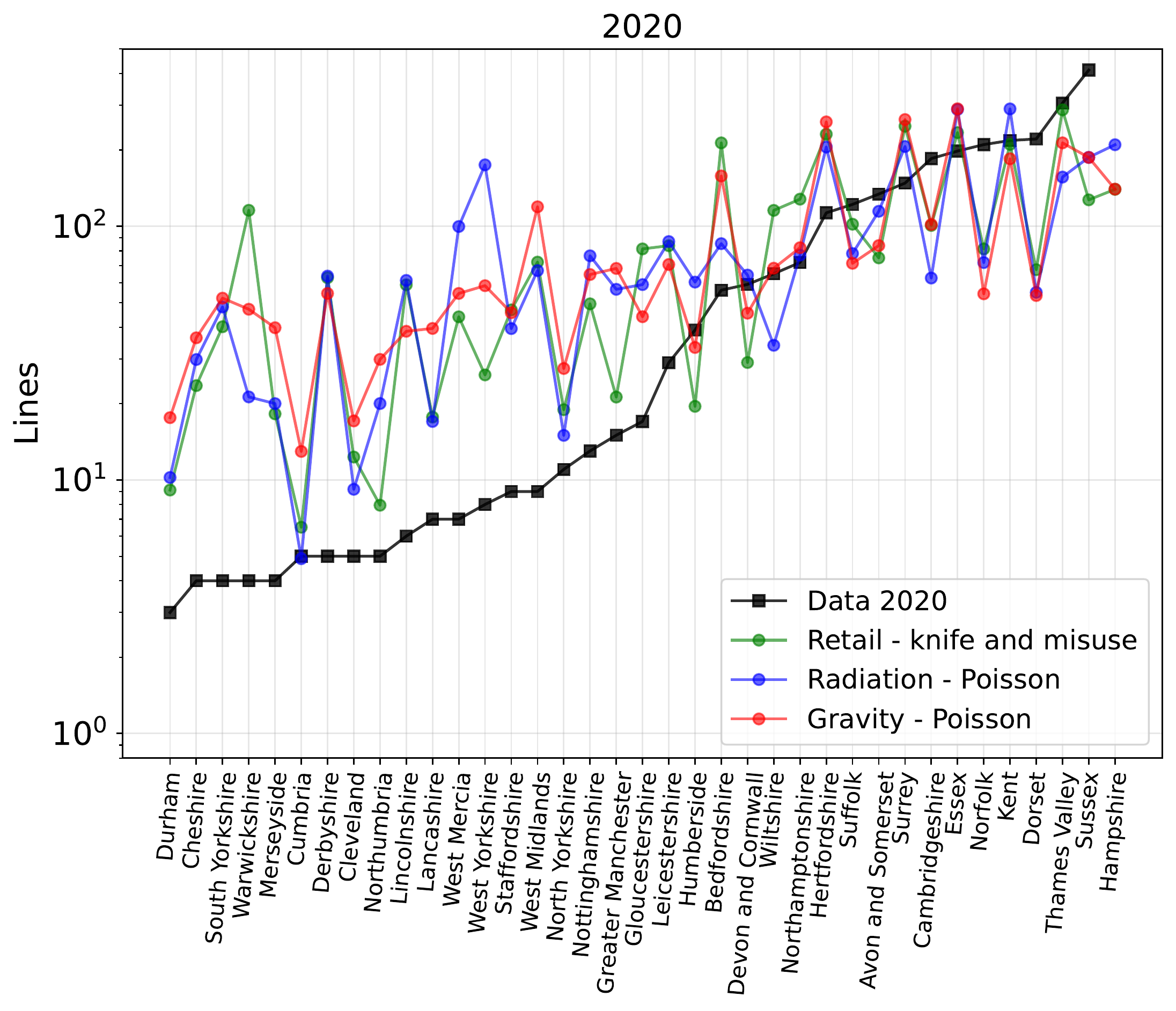}
        \caption{\label{fig:dispersion_20}}
    \end{subfigure}
    \caption{\label{fig:dispersion}Data points and modelled lines ordered by police force for 2019 (a) and 2020 (b). Note: for both years, there were 0 lines detected in Durham. As the plot is in log scale, this data point was not included.}
\end{figure}

Once we have obtained the best performing models, we proceed to compare and analyse the simulated distribution of modelled lines. In Figure \ref{fig:dispersion} we present the different models compared with the Metropolitan Police Data for 2019 and 2020. 

The four best performing models (Retail models) act almost identically, so we only depict models 1, 5 and 6 from Table \ref{tab:results_calibration}. \\

The three models tend to overestimate the detected connections to places with less than ~70 lines, while tending to underestimate them in police territories with more than ~100 lines detected. 


Each one of the models have different ways of understanding the dispersion of flow in a given space. On the one hand, the calibrated Radiation model sees the flow from London to another given police territory as a process of sorting opportunities presented on the way. After our calibration, opportunities here are seen as proportional to the population by the value of $\rho$ given in Table \ref{tab:results_calibration}. Thus, we are actually exploring how the population is distributed in England.

On the other hand, the Retail model understands flow as a balance of with respect to travel times and the other social variables using an exponential distribution. This means that flow from London to another police territory is given by how much time is spent commuting with respect to the other police territories and how much the other benefits/cost relate to it. Thus, a closer place from London would be favoured with respect to a farther one. However, given that this consideration is given by an exponential distribution, we can expect a slow decrease of lines when increasing travel times (light tail distribution).  

Finally, the Gravity model explores the flow with respect to the distance between two places and the population of the target place. In that sense, closer and more populous locations would take most of the outflow, while distant and less populated locations would be disfavoured by the model.\\

\begin{figure}[ht]
    \centering
    \begin{subfigure}[b]{0.24\textwidth}
        \centering
        \includegraphics[width=\textwidth]{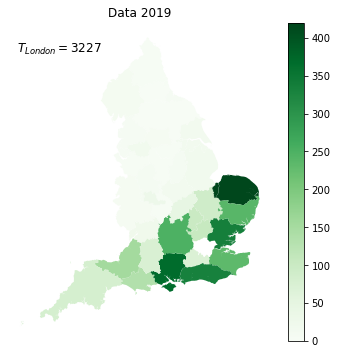}
        \caption{\label{fig:Data2019}}
    \end{subfigure}
    \begin{subfigure}[b]{0.24\textwidth}
        \centering
        \includegraphics[width=\textwidth]{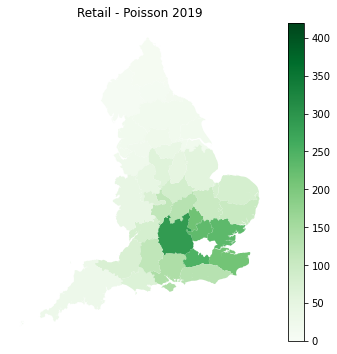}
        \caption{\label{fig:RetailPoisson2019}}
    \end{subfigure}
        \begin{subfigure}[b]{0.24\textwidth}
        \centering
        \includegraphics[width=\textwidth]{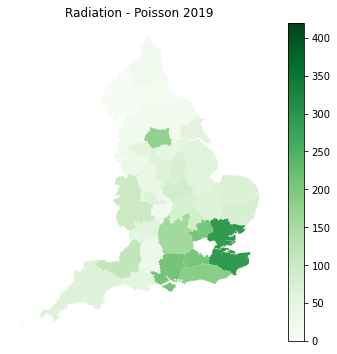}
        \caption{\label{fig:RadiationPoisson2019}}
    \end{subfigure}
    \begin{subfigure}[b]{0.24\textwidth}
        \centering
        \includegraphics[width=\textwidth]{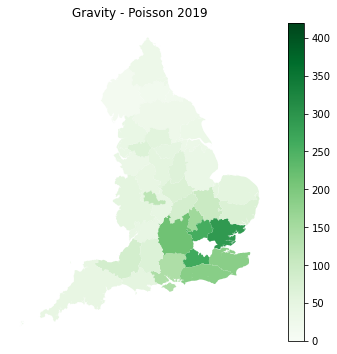}
        \caption{\label{fig:GravityPoisson2019}}
    \end{subfigure}
    \hfill
    \centering
    \begin{subfigure}[b]{0.24\textwidth}
        \centering
        \includegraphics[width=\textwidth]{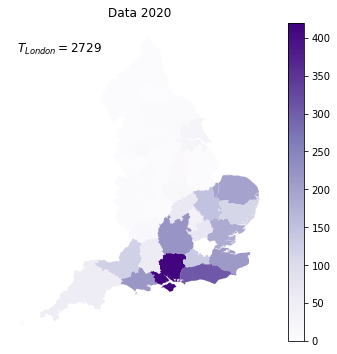}
        \caption{\label{fig:Data2020}}
    \end{subfigure}
    \begin{subfigure}[b]{0.24\textwidth}
        \centering
        \includegraphics[width=\textwidth]{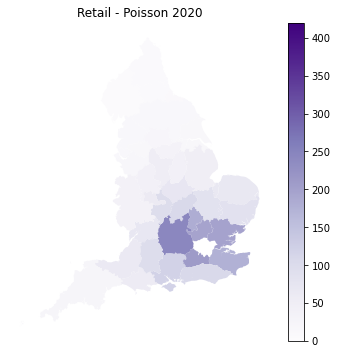}
        \caption{\label{fig:RetailPoisson2020}}
    \end{subfigure}
        \begin{subfigure}[b]{0.24\textwidth}
        \centering
        \includegraphics[width=\textwidth]{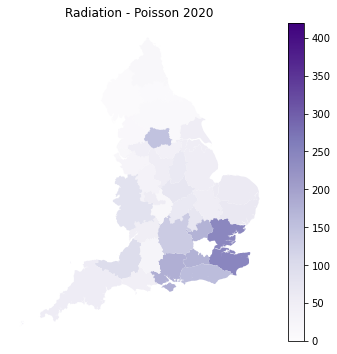}
        \caption{\label{fig:RadiationPoisson2020}}
    \end{subfigure}
    \begin{subfigure}[b]{0.24\textwidth}
        \centering
        \includegraphics[width=\textwidth]{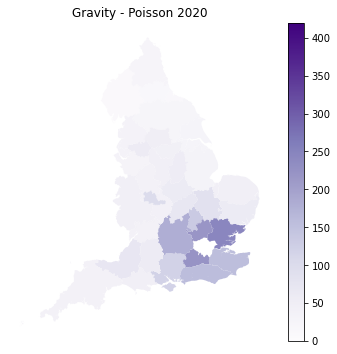}
        \caption{\label{fig:GravityPoisson2020}}
    \end{subfigure}
    \caption{\label{fig:models}Heatmaps for the Metropolitan Police data for 2019 (\ref{fig:Data2019}) and 2020 (\ref{fig:Data2020}), and the three different models tested: Retail (\ref{fig:RetailPoisson2019} and \ref{fig:RetailPoisson2020}), Radiation (\ref{fig:RadiationPoisson2019} and \ref{fig:RadiationPoisson2020}) and Gravity (\ref{fig:GravityPoisson2019} and \ref{fig:GravityPoisson2020}).}
\end{figure}

\begin{figure}[ht]
    \centering    
    \begin{subfigure}[b]{0.32\textwidth}
        \centering
        \includegraphics[width=\textwidth]{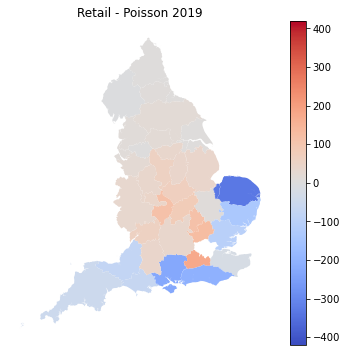}
        \caption{\label{fig:DiffRetailPoisson2019}}
    \end{subfigure}
    \begin{subfigure}[b]{0.32\textwidth}
        \centering
        \includegraphics[width=\textwidth]{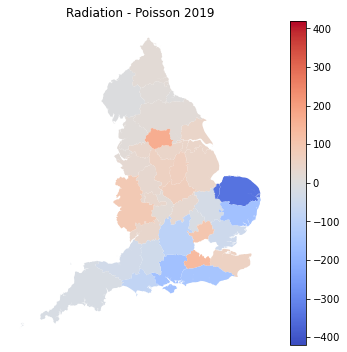}
        \caption{\label{fig:DiffRadiationPoisson2019}}
    \end{subfigure}
    \begin{subfigure}[b]{0.32\textwidth}
        \centering
        \includegraphics[width=\textwidth]{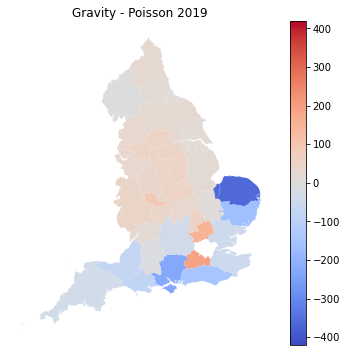}
        \caption{\label{fig:DiffGravityPoisson2019}}
    \end{subfigure}
    \vfill
    \centering    
    \begin{subfigure}[b]{0.32\textwidth}
        \centering
        \includegraphics[width=\textwidth]{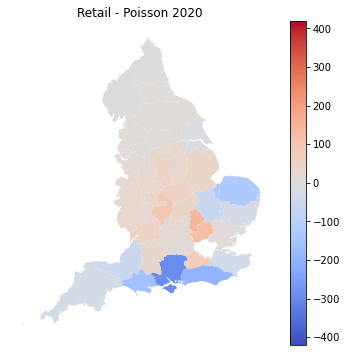}
        \caption{\label{fig:DiffRetailPoisson2020}}
    \end{subfigure}
    \begin{subfigure}[b]{0.32\textwidth}
        \centering
        \includegraphics[width=\textwidth]{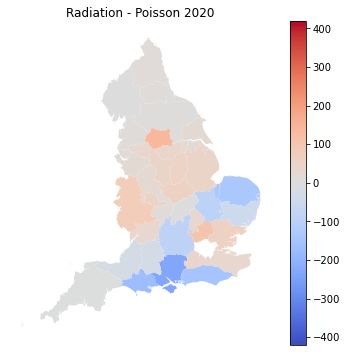}
        \caption{\label{fig:DiffRadiationPoisson2020}}
    \end{subfigure}
    \begin{subfigure}[b]{0.32\textwidth}
        \centering
        \includegraphics[width=\textwidth]{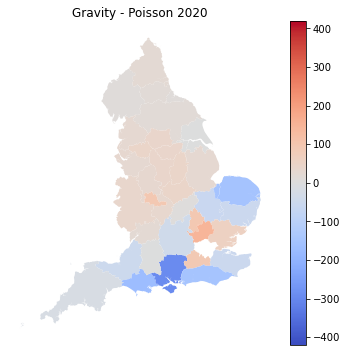}
        \caption{\label{fig:DiffGravityPoisson2020}}
    \end{subfigure}
    \caption{\label{fig:diff}Heatmaps showing the difference between the modelled distribution of lines with respect to the Metropolitan Police data for 2019 and 2020.}
\end{figure}

To further understand the different results shown in Table \ref{tab:results_calibration} and in Figure \ref{fig:dispersion}, we map the different  models and compare them with the Metropolitan Police data. This is shown in Figure \ref{fig:models}. While Figure \ref{fig:Data2019} and \ref{fig:Data2020} present the Metropolitan Police data for 2019 and 2020, the rest present the modelled spatial distribution of lines.
We also present the differences between the Metropolitan Police data and the models in Figure \ref{fig:diff}. Red zones correspond to territories overestimated by the model, while blue zones correspond to territories underestimated by the model. \\

We start by analysing Figures \ref{fig:Data2019} and \ref{fig:Data2020} corresponding to the Metropolitan Police data. The first thing to notice is the decrease of detected lines in 2020 with respect to 2019. This effect can be given by mainly two factors taking into account the COVID-19 situation throughout 2020: the police had a smaller capacity to detect, or indeed the reduced mobility in the country reduced the number of connections. However, the decrease is not generalised and we can observe an increase in some police territories from 2019 to 2020, as in Hampshire (South of England) where we find the maximum number for 2020. 

An important second element to note from the ground truth data is a very high share of the total lines (94.02\% for 2019 and 93.77\% for 2020) concentrated in 16 out of the 37 police territories considered. This set of 16 police territories, in addition to London, is considered to be the ``South'' of England, a social region with no administrative recognition which encloses the most developed parts of England and which opposes the ``North'' of England, where more industrial cities like Manchester and Liverpool are found (for a study using percolation theory please refer to \cite{Arcaute2016}).  \\

The ``North-South'' division is an element which none of the models captured. However, we can still see different ways of simulating the problem in Figure \ref{fig:models}.
As discussed before, the Retail model distributes the lines in what appears a concentric fashion with respect to London, leaning towards the centre of England. This can be seen more clearly in Figures \ref{fig:DiffRetailPoisson2019} and \ref{fig:DiffRetailPoisson2020}, where we observe an overestimation in the Midlands and an underestimation of the coastal territories of the ``South''.  Note how the far South West of England (Cornwall and Devon), which is farther away in travel times than the centre of England from London, is underrepresented. This fact accounts for an argument in which the operators in London would not have as primordial element for establishing connections the travel times to the different territories. This argument is supported by the opposite fact, where we observe an overestimation by the retail-gravity model in more connected places from London, like the West Midlands (Birmingham) and Warwickshire (south of Birmingham).

The radiation model understands the flow in a different fashion, as seen in Figures \ref{fig:RadiationPoisson2019} and \ref{fig:RadiationPoisson2020}. In a similar way as the Retail model, the ring surrounding London is still catching an important number of lines. However, we can also observe a number of relatively large hotspots, particularly in West Yorkshire (North of England) and in West Mercia (border with Wales). While the former territory includes important cities and urban centres such as Leeds and Bradford, West Mercia is a diverse territory with dense suburban counties belonging to the Birmingham metropolitan area and more rural areas towards Wales, like Shropshire. In Figures \ref{fig:DiffRadiationPoisson2019} and \ref{fig:DiffRadiationPoisson2020} we observe also how the territories between West Yorkshire and London were filled with lines by the Radiation model. It is also important to note how the big metropolitan areas in England such as Birmingham do not appear as hotspots in Figures \ref{fig:RadiationPoisson2019} and \ref{fig:RadiationPoisson2020}. 
  
Both models described above tend to distribute the number of lines in the centre of England, while avoiding the big cities. This is in contrast with the Gravity model (Figures \ref{fig:GravityPoisson2019} and \ref{fig:GravityPoisson2020}) where we observe the appearance of Birmingham and Manchester (2nd and 3rd most populous cities in the UK) as county lines hotspots. 

The three models do not detect the territories where the maximum number of lines are detected, like Norfolk in 2019 and Hampshire in 2020. On one hand this is a sign of no overfitting from both models, but on the other hand makes very difficult for the models to detect future hotspots in the South of England.

\section{Discussion}
\label{sec:discussion}

In the present work we study the County Lines Model (CLM) distribution method of illicit drugs in England. Our aim is to shed some light around the territorial logic behind the data accounting the detected connections (lines) by the Metropolitan Police of London in other police territories \cite{Met2019, Met2020, Black2020_evidence, Black2020_summary}. 

We understand the number of detected lines as a flow of people/merchandise that starts in London and finishes in a given police territory. In that sense, by modelling the flow from one place to another and compare it with the available data, we obtain information about which elements are present when establishing a local market. 

Three different models are studied and compared. Each one of them follows, by construction, different logics about how to understand flow from one place to another. The first one, the Gravity model \cite{Noulas2012, Anderson2010}, sees flow as proportional to the population of both places, while inversely proportional to the distance between them. We take this model as our benchmark as represents the classic idea that populous places would draw more attention than others given the same distance. The second, the Radiation model \cite{Simini2012}, understands flow to a given place as a process of sorting the opportunities before arriving to the final destination. With this second model we tested if the distribution of the population (in comparison to a single population spot) in England was involved in the decision making. Finally, the Retail model \cite{Wilson2006, Wilson2008} takes into account the balance between the benefits and costs of establishing a flow towards a particular place. This final model allows to include as potential benefits/costs different social variables that we tested, like police workforce, knife crime events, hospital admissions for drug poisoning and for drug misuse. 

We train the models using the Metropolitan Police data, and compare them using the Bayesian Information Criterion \cite{Altmann2020}, and the S\o{}rensen-Dice index \cite{Piovani2018} over a cross-validation. We also test two different loss functions, the classic Mean-square error and the one derived from a Poissonian likelihood.

The best performing model is the Retail model with different combinations of social variables and trained with the 2019 data and the Poisson loss function. We find that for certain combinations of social variables, the Retail model would give better results than others. Indeed, the hospital admissions by misuse of drugs \textit{per capita} and knife crime events \textit{per capita} are the two most influential variables to obtain better results. In particular, knife crimes shows to be more important to hospital admissions when compared one to one. 

The Radiation and Gravity model also perform correctly when trained with the 2019 data and the Poisson loss function. However, when comparing the geographical distribution of both model to the Metropolitan Police data, these two models predict hotspots in populous regions of England where no important number of lines was detected by the Met. Police.  

According to our ground truth, the distribution of the great majority of lines (~93\%) is over 16 of the 37 police territories in England, which form the union of the South West, the South East and the East of England. This territory is known as the ``South'' of England. 

While the Gravity and Radiation model overestimated different territories outside the South of England with large populations, the Retail model did it in a more diffused way. This is due to the exponential for of the model.

None of the three models could capture the hard border that the data shows between the South of England and the rest of the country. This raises the question about the characteristics in the 16 police territories that represent the 'South' of England that make them so attractive to CL operators. A first hypothesis is that the CLM, although not reported in literature, actually acts within a more organised structure which can restrict itself to distribute in a given territory, as seen for other criminal organisations. In other words, even though not mentioned in the public information by the UK government, different CLM gangs operating from London could restrict themselves to these 16 police territories as a measure to do not enter in an open conflict with other gangs from other CLM hubs. This hypothesis could be studied by having data from the detected lines from the other important CLM hubs, like Birmingham, Liverpool, Manchester and West Yorkshire. In that sense, we could expect a localised distribution in the 'North' of England, obtaining then a polycentric structure within the territory. 

However, if data from other CLM hubs would not comply with the segregation and rather concentrate in a subset of the 16 police territories considered as the ``South'', then we would have a particularity of the consumers in those areas. This would also be of interest, as the population in this subset would have to have a distinction with respect to the other big metropolitan and rural areas of the 21 police territories left. This distinction, although might be related to a particularity of the consumers, would also address the findings already obtained before in quantitative studies \cite{Arcaute2016}, where a clear distinction between the urban network between the South and the rest of England was found using percolation analysis. 

The hypothesis about a polycentric structure could be supported by our findings on how knife crime events and hospital admissions by misuse of drugs are a cost to line operators. The fact that knife crime events appear as a cost might point to an avoidance from the operators to certain gangs so conflict is spared. Hospital admissions, on the other hand, are used as a proxy to illicit drugs consumers given the lack of public information about it. In that sense, the fact that the hospital admissions variable is one of the two most influential variables, combined with the knife crime variable, could be interpreted as county line operators avoiding places where there already is enough competition for them to handle. This competition can be regarded as possible origin of conflicts (knife crimes) and responsible of having a greater share of the illicit drug consumption market in a given territory. \\

We demonstrate that the logic behind the county line operators is not as simple as an offer-demand one \cite{Met2019, Met2020, Black2020_evidence, Black2020_summary, WestMidlands2020, NCA2018, NCLCC2021}, but actually might follow a social structure of the country, while also avoiding conflict with other gangs and markets already filled with competition. This by itself can be of great help for law-enforcement bodies, as it gives a good lead on where to look for the presence of county lines from London: places within the 16 police territories where there is not an important number of knife crimes \textit{per capita}. There is no mention of these factors in the reviewed literature. 

This works also allows to implement a better coordination between local police forces, as the Metropolitan Police of London would only need to coordinate with 43\% of the English police forces to tackle the 93\% of the lines detected.

The main limitation of this work is the lack of data. Having a larger dataset both in the temporal dimension and the territorial origin could make us obtain a more complete analysis of the county lines model not only in England but in Great Britain as a whole (no county lines has been identified in Northern Ireland).
By obtaining data from other police territories like Liverpool (Merseyside police), Manchester, Birmingham (West Midlands Police) or West Yorkshire we could, as possible extension of this work, to analyse the hypothesis described above.

\bibliographystyle{unsrt}
\bibliography{references}

\begin{thebibliography}{10}

\bibitem{NCA2018}
National {C}rime {A}gency.
\newblock {\em Intelligence {A}ssessment: {C}ounty {L}ines {D}rug {S}upply,
  {V}ulnerability and {H}arm 2018}, January 2019.

\bibitem{Black2020_evidence}
Carol Black.
\newblock {\em Review of Drugs: evidence relating to drug use, supply and
  effects, including current trends and future risks.}
\newblock Home Office, February 2020.

\bibitem{Andell2018}
Paul Andell and John Pitts.
\newblock The end of the line? the impact of county lines drug distribution on
  youth crime in a target destination.
\newblock {\em Youth \& Policy}, January 2018.

\bibitem{Robinson2019}
Grace Robinson.
\newblock Working county lines: Child criminal exploitation and illicit drug
  dealing in {G}lasgow and {M}erseyside.
\newblock {\em International Journal of Offender Therapy \& Comparative
  Criminology}, 63(5):694--712, April 2019.

\bibitem{Stone2018}
Nigel Stone.
\newblock Child criminal exploitation: ‘county lines’, trafficking and
  cuckooing.
\newblock {\em Youth Justice}, 18(3):285--293, 2018.

\bibitem{CAMBER2020}
Rebecca Camber.
\newblock 1,500 more county lines drugs gangs in just a year.
\newblock {\em Daily Mail}, pages 40--41, April 2020.

\bibitem{Moyle2019}
Leah Moyle, Andrew Childs, Ross Coomber, and Monica~J Barratt.
\newblock Drugsforsale: An exploration of the use of social media and encrypted
  messaging apps to supply and access drugs.
\newblock {\em International Journal of Drug Policy}, 63:101--110, 2019.

\bibitem{Coombes2018}
Rebecca Coombes.
\newblock Brexit: the clock is ticking.
\newblock {\em The BMJ}, 362:k4057, 2018.

\bibitem{NCLCC2021}
NCLCC Silver and Bronze Intelligence.
\newblock Nclcc county lines strategic assessment 2020/2021.
\newblock Technical report, National County Lines Coordination Centre, 2021.

\bibitem{Black2020_summary}
Carol Black.
\newblock {\em Review of Drugs. Executive Summary.}
\newblock Home Office, February 2020.

\bibitem{Anderson2010}
James~E Anderson.
\newblock The gravity model.
\newblock Working Paper 16576, National Bureau of Economic Research, December
  2010.

\bibitem{Simini2012}
Filippo Simini, Marta~C. González, Amos Maritan, and Albert-László
  Barabási.
\newblock A universal model for mobility and migration patterns.
\newblock {\em Nature}, 484:96--100, 2012.

\bibitem{Wilson2008}
Alan Wilson.
\newblock Boltzmann, {L}otka and {V}olterra and spatial structural evolution:
  an integrated methodology for some dynamical systems.
\newblock {\em Journal of The Royal Society Interface}, 5(25):865--871, 2008.

\bibitem{Piovani2018}
Duccio Piovani, Elsa Arcaute, Gabriela Uchoa, Alan Wilson, and Michael Batty.
\newblock Measuring accessibility using gravity and radiation models.
\newblock {\em Royal Society Open Science}, 5(9):171668, 2018.

\bibitem{Commons2019}
Alexaner Bellis, Grahame Allen, and Lukas Audickas.
\newblock Knife crime statistics.
\newblock Technical report, House of {c}ommons {L}ibrary, December 2019.

\bibitem{Met2019}
Rescue and Response~Project Analysts.
\newblock Rescue and response county lines project. supporting young
  {L}ondoners affected by county lines exploitation., 2019.

\bibitem{Met2020}
Rescue and Response~Project Analysts.
\newblock Rescue and response county lines project. supporting young
  {L}ondoners affected by county lines exploitation., 2010.

\bibitem{WestMidlands2020}
{S}haer Supt~{M}at.
\newblock County {L}ines, 2020.

\bibitem{Commons2018}
Sarah Pepin.
\newblock County lines exploitation in {L}ondon.
\newblock Technical report, House of {C}ommons {L}ibrary, 2018.

\bibitem{Gavin2018}
Gavin Madeley.
\newblock County lines gangs using children to shift drugs in scotland.
\newblock {\em Daily Mail}, pages 18--20, October 2018.

\bibitem{Ross2017}
Ross Coomber and Leah Moyle.
\newblock {The Changing Shape of Street-Level Heroin and Crack Supply in
  England: Commuting, Holidaying and Cuckooing Drug Dealers Across ‘County
  Lines’}.
\newblock {\em The British Journal of Criminology}, 58(6):1323--1342, 11 2017.

\bibitem{Roman-Urrestarazuk4003}
Andres Roman-Urrestarazu, Roy Robertson, Justin Yang, Alison McCallum,
  Christina Gray, Martin McKee, and John Middleton.
\newblock European monitoring centre for drugs and drug addiction has a vital
  role in the {U}{K}{\textquoteright}s ability to respond to illicit drugs and
  organised crime.
\newblock {\em The BMJ}, 362, 2018.

\bibitem{Wilson2006}
Alan~G Wilson.
\newblock Ecological and urban systems models: Some explorations of
  similarities in the context of complexity theory.
\newblock {\em Environment and Planning A: Economy and Space}, 38(4):633--646,
  2006.

\bibitem{Davies2013}
Toby~P. Davies, Hannah~M. Fry, Alan~G. Wilson, and Steven~R. Bishop.
\newblock A mathematical model of the {L}ondon riots and their policing.
\newblock {\em Scientific Reports}, 3:2045--2322, 2013.

\bibitem{Noulas2012}
Anastasios Noulas, Salvatore Scellato, Renaud Lambiotte, Massimiliano Pontil,
  and Cecilia Mascolo.
\newblock A tale of many cities: Universal patterns in human urban mobility.
\newblock {\em PLOS ONE}, 7(5):1--10, 05 2012.

\bibitem{Piovani2017}
Duccio Piovani, Carlos Molinero, and Alan Wilson.
\newblock Urban retail location: Insights from percolation theory and spatial
  interaction modeling.
\newblock {\em PLOS ONE}, 12(10):1--13, 10 2017.

\bibitem{Masucci2013}
A.~Paolo Masucci, Joan Serras, Anders Johansson, and Michael Batty.
\newblock Gravity versus radiation models: On the importance of scale and
  heterogeneity in commuting flows.
\newblock {\em Phys. Rev. E}, 88:022812, Aug 2013.

\bibitem{Yang2014}
Yingxiang Yang, Carlos Herrera, Nathan Eagle, and Marta~C. González.
\newblock Limits of predictability in commuting flows in the absence of data
  for calibration.
\newblock {\em Scientific Reports}, 4:5662, 2014.

\bibitem{Altmann2020}
Eduardo~G. Altmann.
\newblock Spatial interactions in urban scaling laws.
\newblock {\em PLOS ONE}, 15(12):1--12, 12 2020.

\bibitem{NHS2019}
NHS~Digital Lifestyles~Team.
\newblock Statistics on drug misuse.
\newblock Technical report, NHS, August 2019.

\bibitem{HomeOfficeWorkforce2019}
John Flatley.
\newblock Police workforce england and wales statistics.
\newblock Technical report, Home Office, August 2019.

\bibitem{GDHI2021}
Trevor Fenton.
\newblock Regional gross disposable household income, uk: 1997 to 2019.
\newblock Technical report, Office for National Statistics, October 2021.

\bibitem{Arcaute2016}
Elsa Arcaute, Carlos Molinero, Erez Hatna, Roberto Murcio, Camilo Vargas-Ruiz,
  A.~Paolo Masucci, and Michael Batty.
\newblock Cities and regions in britain through hierarchical percolation.
\newblock {\em Royal Society Open Science}, 3(4):150691, 2016.

\bibitem{ONSDrugs2019}
Emyr John.
\newblock Drug-related deaths by local authority, england and wales.
\newblock Technical report, ONS, August 2019.

\end{thebibliography}

\begin{appendices}
\section{Table of models}
\label{sec:appendix_a}

In Table \ref{tab:models} we present all the different models tested. They are numbered as shown in Figure \ref{fig:BIC} and Figure \ref{fig:Sorensen}.


\begin{longtable}{c|c|c|l}
    \caption{\label{tab:models}List of all trained models.} \\
      & Model & Loss Function & Free parameters calibrated\\
     \hline
     1 & Gravity & MSE & $b$, $c$ \\
     2 & Gravity & Poisson & $b$, $c$ \\
     3 & Radiation & MSE & $\rho$, $r$ \\
     4 & Radiation & Poisson & $\rho$, $r$ \\ 
     5 & Retail & Poisson & $\beta$ (travel times) \\
     6 & Retail & Poisson & $\beta$, $\alpha_1$  (hospital admissions by misuse of drugs)\\ 
     7 & Retail & Poisson & $\beta$, $\alpha_2$ (hospital admissions by poisoning of drugs)\\
     8 & Retail & Poisson & $\beta$, $\alpha_3$ (police workforce) \\
     9 & Retail & Poisson & $\beta$, $\alpha_4$ (knife crime events) \\
     10 & Retail & Poisson & $\beta$, $\alpha_5$ (gross dispensable household income) \\
     11 & Retail & Poisson & $\beta$, $\alpha_1$, $\alpha_2$ \\
     12 & Retail & Poisson & $\beta$, $\alpha_1$, $\alpha_3$ \\
     13 & Retail & Poisson & $\beta$, $\alpha_1$, $\alpha_4$ \\
     14 & Retail & Poisson & $\beta$, $\alpha_1$, $\alpha_5$ \\
     15 & Retail & Poisson & $\beta$, $\alpha_2$, $\alpha_3$ \\
     16 & Retail & Poisson & $\beta$, $\alpha_2$, $\alpha_4$ \\
     17 & Retail & Poisson & $\beta$, $\alpha_2$, $\alpha_5$ \\
     18 & Retail & Poisson & $\beta$, $\alpha_3$, $\alpha_4$ \\
     19 & Retail & Poisson & $\beta$, $\alpha_3$, $\alpha_5$ \\
     20 & Retail & Poisson & $\beta$, $\alpha_4$, $\alpha_5$ \\
     21 & Retail & Poisson & $\beta$, $\alpha_1$, $\alpha_2$, $\alpha_3$ \\
     22 & Retail & Poisson & $\beta$, $\alpha_1$, $\alpha_2$, $\alpha_4$ \\
     23 & Retail & Poisson & $\beta$, $\alpha_1$, $\alpha_2$, $\alpha_5$ \\
     24 & Retail & Poisson & $\beta$, $\alpha_1$, $\alpha_3$, $\alpha_4$ \\
     25 & Retail & Poisson & $\beta$, $\alpha_1$, $\alpha_3$, $\alpha_5$ \\
     26 & Retail & Poisson & $\beta$, $\alpha_2$, $\alpha_3$, $\alpha_4$ \\
     27 & Retail & Poisson & $\beta$, $\alpha_2$, $\alpha_3$, $\alpha_5$ \\
     28 & Retail & Poisson & $\beta$, $\alpha_3$, $\alpha_4$, $\alpha_5$ \\    
     29 & Retail & Poisson & $\beta$, $\alpha_2$, $\alpha_4$, $\alpha_5$ \\
     30 & Retail & Poisson & $\beta$, $\alpha_1$, $\alpha_4$, $\alpha_5$ \\
     31 & Retail & Poisson & $\beta$, $\alpha_1$, $\alpha_2$, $\alpha_3$, $\alpha_4$ \\
     32 & Retail & Poisson & $\beta$, $\alpha_1$, $\alpha_2$, $\alpha_3$, $\alpha_5$ \\
     33 & Retail & Poisson & $\beta$, $\alpha_1$, $\alpha_2$, $\alpha_4$, $\alpha_5$ \\
     34 & Retail & Poisson & $\beta$, $\alpha_1$, $\alpha_3$, $\alpha_4$, $\alpha_5$ \\
     35 & Retail & Poisson & $\beta$, $\alpha_2$, $\alpha_3$, $\alpha_4$, $\alpha_5$ \\
     36 & Retail & Poisson & $\beta$, $\alpha_1$, $\alpha_2$, $\alpha_3$, $\alpha_4$, $\alpha_5$ \\
     37 & Retail & MSE & $\beta$ \\
     38 & Retail & MSE & $\beta$, $\alpha_1$ \\ 
     39 & Retail & MSE & $\beta$, $\alpha_2$ \\
     40 & Retail & MSE & $\beta$, $\alpha_3$ \\
     41 & Retail & MSE & $\beta$, $\alpha_4$ \\
     42 & Retail & MSE & $\beta$, $\alpha_5$ \\
     43 & Retail & MSE & $\beta$, $\alpha_1$, $\alpha_2$ \\
     44 & Retail & MSE & $\beta$, $\alpha_1$, $\alpha_3$ \\
     45 & Retail & MSE & $\beta$, $\alpha_1$, $\alpha_4$ \\
     46 & Retail & MSE & $\beta$, $\alpha_1$, $\alpha_5$ \\
     47 & Retail & MSE & $\beta$, $\alpha_2$, $\alpha_3$ \\
     48 & Retail & MSE & $\beta$, $\alpha_2$, $\alpha_4$ \\
     49 & Retail & MSE & $\beta$, $\alpha_2$, $\alpha_5$ \\
     50 & Retail & MSE & $\beta$, $\alpha_3$, $\alpha_4$ \\
     51 & Retail & MSE & $\beta$, $\alpha_3$, $\alpha_5$ \\
     52 & Retail & MSE & $\beta$, $\alpha_4$, $\alpha_5$ \\
     53 & Retail & MSE & $\beta$, $\alpha_1$, $\alpha_2$, $\alpha_3$ \\
     54 & Retail & MSE & $\beta$, $\alpha_1$, $\alpha_2$, $\alpha_4$ \\
     55 & Retail & MSE & $\beta$, $\alpha_1$, $\alpha_2$, $\alpha_5$ \\
     56 & Retail & MSE & $\beta$, $\alpha_1$, $\alpha_3$, $\alpha_4$ \\
     57 & Retail & MSE & $\beta$, $\alpha_1$, $\alpha_3$, $\alpha_5$ \\
     58 & Retail & MSE & $\beta$, $\alpha_2$, $\alpha_3$, $\alpha_4$ \\
     59 & Retail & MSE & $\beta$, $\alpha_2$, $\alpha_3$, $\alpha_5$ \\
     60 & Retail & MSE & $\beta$, $\alpha_3$, $\alpha_4$, $\alpha_5$ \\    
     61 & Retail & MSE & $\beta$, $\alpha_2$, $\alpha_4$, $\alpha_5$ \\
     62 & Retail & MSE & $\beta$, $\alpha_1$, $\alpha_4$, $\alpha_5$ \\
     63 & Retail & MSE & $\beta$, $\alpha_1$, $\alpha_2$, $\alpha_3$, $\alpha_4$ \\
     64 & Retail & MSE & $\beta$, $\alpha_1$, $\alpha_2$, $\alpha_3$, $\alpha_5$ \\
     65 & Retail & MSE & $\beta$, $\alpha_1$, $\alpha_2$, $\alpha_4$, $\alpha_5$ \\
     66 & Retail & MSE & $\beta$, $\alpha_1$, $\alpha_3$, $\alpha_4$, $\alpha_5$ \\
     67 & Retail & MSE & $\beta$, $\alpha_2$, $\alpha_3$, $\alpha_4$, $\alpha_5$ \\
     68 & Retail & MSE & $\beta$, $\alpha_1$, $\alpha_2$, $\alpha_3$, $\alpha_4$, $\alpha_5$ \\
\end{longtable}
\section{Database}
\label{sec:supplementary_matieral}

|
The present appendix is divided in two sections. In Section \ref{ssec:database} we present the different elements worth mentioning from the database. These include the different data that are included, their respective resolution, format and sources. 

We also include the different territorial divisions used in Section \ref{ssec:territorial_resolution}. These include the territorial divisions for NHS Local Authorities, Local Police Forces and English regions. 

\section{The database}
\label{ssec:database}

All the compiled information is obtained from different British Governmental websites, being public, open and shared with the \href{http://www.nationalarchives.gov.uk/doc/open-government-licence/version/3/}{Open Government Licence}. Only the used travel time matrices are obtained from a third party company, which is the Maps API from Google®.

Most of the data is presented in \verb|.csv| format and thought to be managed as \textit{data frames} objects with packages as \verb|pandas| for Python or \verb|data.table| for R. \\

Most of the data is presented as time series. Coming from the British Government, the year steps from one data to another are not from January to December, but rather from April to March. This is because the Government takes the fiscal year as unit of time. In that sense, when a measurement reads for a ``2012'', this actually means that the measurement refers to the fiscal year starting in April 1, 2011 and finishes on March 31, 2012. 

In the following, we present each kind of data used in this project. Subsections of missing data are also added. This is only to acknowledge the raw data that has not been fully processed.

\subsection{Drug related hospital admissions data}

NHS hospitals present annually data about their hospital admissions. The particular set of hospital admissions related to drugs comprise three different types:
\begin{enumerate}
    \item NHS hospital finished admissions where there was a \textbf{primary diagnosis of drug related mental health and behavioural disorders}.
    \item NHS hospital finished admission episodes with a \textbf{primary or secondary diagnosis of drug related mental and behavioural disorders}.
    \item NHS hospital finished admissions where a \textbf{primary diagnosis of poisoning by drugs}.
\end{enumerate}

We include the time series for each kind of hospital admission from 2009 to 2019 for different geographical resolutions: England, its 9 regions, 39 police force areas and 131 counties described in Section \ref{ssec:territorial_resolution}. We also include their respective time series for admissions by 100 thousand inhabitants. Depending of the resolution, the normalisation is done using the population of the territorial unit. That is, the measure for Northumbria in 2012 is done with the population of Northumbria in 2012. 

At national level we include the age distribution for each kind of admission. We also include the distribution of diagnoses for admissions type 1 and 3. However, for admissions type 3 these were published only from 2013. \\

Admissions type 2 are not considered into the analysis and are only considered for reference. This decision is based on the lack of information from the primary and secondary diagnoses, thus being unable to detect the underlying causes of the admissions. \\

The main source is the \href{https://digital.nhs.uk/data-and-information/publications/statistical/statistics-on-drug-misuse}{Statistics on Drug Misuse} published by NHS Digital annually.

\subsection{Drug related deaths data}

Statistics about drug related deaths in England and Wales are published each year. In this case, data has an inherent delay caused by the difference between the decease date and the registered date. The delay in England for 2018 had a median of 181 days, and a median delay of 172 days for 2017 according to the publishers \cite{ONSDrugs2019}. This makes the interpretation from this statistic more difficult to handle, as some deaths registered in a particular year could have happened more than a year before.

The publishers also make the distinction between the deaths caused by poisoning of drugs and those deaths by poisoning of which were caused by misuse of drugs. This is a subtle distinction, as the drug poisoning death is defined by the WHO's \href{https://www.who.int/classifications/icd/en/}{International Classification of Diseases}. The drug misuse death is a drug poisoning death which also involves a drug abuse or dependence. \\

We publish the time series (2009-2018) for the different resolutions handled: England, its 9 regions, the 39 local police areas and the 131 counties described in Section \ref{ssec:territorial_resolution}. We also include for each one of these resolutions their respective time series for deaths by 100 thousand inhabitants. Depending of the resolution, the population of the territorial unit is used. That means that for the number of deaths for each 100 thousand inhabitants in Essex for 2012, the population for Essex in 2012 is used.

In the bottom three resolutions (local authorities, police forces and regions) we only present the total number of deaths by poisoning and by misuse. However, at the national (England) level we also present the time series for underlying causes, age distribution for deaths by misuse and by poisoning, and the age distribution by drug for the total number of deaths. \\

Given that the data does not cover 2019 and 2020, this data was not used for the analysis of this work. \\

The main source of the deaths related to drugs data is the \href{https://www.ons.gov.uk/peoplepopulationandcommunity/birthsdeathsandmarriages/deaths/bulletins/deathsrelatedtodrugpoisoninginenglandandwales/2018registrations#things-you-need-to-know-about-this-release}{Deaths related to drug poisoning in England and Wales: 2018 registrations} published annually by the Office for National Statistics. 

\subsection{Number of hospital beds data}

The number of hospital beds was collected for the three different resolutions. However, we only recommend data for the regional and local police resolutions. This is due to the fact that an important number of reported hospital beds are an addition for different hospitals in different local authorities. An example of this is the Guy's and St. Thomas' Hospitals: the hospital beds are reported as an addition for both hospitals, while one is in the London Borough of Lambeth, and the other in the London Borough of Southwark. This of course is solved when counting the hospital beds for the Metropolitan Police resolution, including most of the Greater London boroughs. 

This data was used to normalise the hospital admissions.

The main data is the NHS database of  \href{https://www.england.nhs.uk/statistics/statistical-work-areas/bed-availability-and-occupancy/bed-data-day-only/}{hospital beds availability. }

\subsection{Police workforce data}

The police workforce data is the only one that is published more than once a year, being published each semester. This allows to know how the workforce varies along each year. The Home Office publishes this data in different ways, considering the number of heads working full-time and part-time jobs, and doing a conversion to the equivalent of heads working only full-time. In all years they include police officers and police staff. However, from 2012 the Home Office includes the numbers for different job titles working in the police workforces. These include community support officers, designated officers and traffic warden. In order to have a coherent database, we only consider the regular officers in the conversion to heads working only full-time.

The time series are presented for data from 2009 to 2019. They are presented for England, its 9 regions and the 39 local police forces, including the British Transport Police and the Central Service Secondments. More information about the different resolutions is found in Section \ref{ssec:territorial_resolution}. For each resolution, workforces are given by annual mean with its correspondent standard deviation (for measurement, we take the workforce number at the beginning, middle and end of each year). We also include the same numbers for each 100k inhabitants. The normalisation is done using the respective population resolution. That means that the mean for each 100k inhabitants for Oxfordshire in 2012 is obtained using the population of the same county in 2012. In the case of the British Transport Police and the Central Service Secondments, the population of England is used. \\

The main source for the workforce data is the \href{https://www.gov.uk/government/collections/police-workforce-england-and-wales}{Police workforce England and Wales statistics} published by the Home Office. 

\subsection{Police numbers of drug seizures data}

The Home office publishes once a year the number of seizures and total quantities by drug and police force. This allows to compile a set of time series (2010-2019) for different drugs at different resolutions (England, regions and police forces).  Also, at England \& Wales resolution we obtain the number of seizures by weight/dosage for different drugs. The available drugs (with their dosage unit) are:
\begin{itemize}
    \item \textbf{Class A drugs}: Cocaine (kg), Crack (kg), Ecstasy (doses), Heroin (kg), LSD (doses), Methadone (doses), Morphine (doses).
    \item \textbf{Class B drugs}: Herbal Cannabis (kg), Cannabis Resin (kg), Cannabis Plants (plants), Amphetamines (kg), Barbiturates (doses), Ketamine* (kg).
    \item \textbf{Class C drugs}: Anabolic steroids (kg), Benzodiazephines (doses), GHB (doses), Temazepam (doses).
\end{itemize}
The main source is the \href{https://www.gov.uk/government/collections/seizures-of-drugs-in-england-and-wales}{Seizure of drugs in England and Wales statistics} published by the Home Office annually.  

Given the lack of data for 2020 and the geographical resolution used, we did not use this data for the analysis of this work.

*: In the fiscal year 2014-2015, Ketamine was reclassified a Class B drug instead of a Class C drug. 
\subsection{Knife crime related data}

Knife crime has been reported by the \href{https://commonslibrary.parliament.uk/research-briefings/sn04304/}{House of Commons library} since 2009. The data is at the police forces resolution. 

\subsection{Disposable Income data}

The Gross Disposable Household Income data is reported the \href{https://www.ons.gov.uk/economy/regionalaccounts/grossdisposablehouseholdincome/bulletins/regionalgrossdisposablehouseholdincomegdhi/1997to2019#gross-disposable-household-income-data}{ONS}. The data is available at county level. We aggregated the income to a police territory resolution using a weighted average using the population of each county. 


\subsection{Demographic data}

As demographic data we include the time series of the population for different resolutions of England from 2009 to 2019. The different resolutions are those described in Section \ref{ssec:territorial_resolution}, and refer to England, the 9 regions conforming England, the English Local Police Forces and the 131 Counties adopted for this project. \\

The main source of the demographic data is the \href{https://www.ons.gov.uk/peoplepopulationandcommunity/populationandmigration/populationestimates/datasets/populationestimatesforukenglandandwalesscotlandandnorthernireland}{Estimate of the population for the UK, England, Wales, Scotland and Northern Ireland}, from the Office for National Statistics. The Estimates are released each year. 

\subsection{Geographic data}

Geographic data is analogous to demographic data. We include different \verb|.geojson| files containing the geometries for Great Britain in different resolutions. The different resolutions are those described in Section \ref{ssec:territorial_resolution}, and refer to England, the 9 regions conforming England, the English Local Police Forces and the 131 counties adopted for this project. We also include a fourth file for Scotland and Wales at local authority resolution. \\

The main source of the geographic data is the \href{https://geoportal.statistics.gov.uk/search?collection=Document&sort=name&tags=all(MAP_ADM)}{Open Geography Portal} from the Office for National Statistics. 

\subsection{Territorial resolutions}
\label{ssec:territorial_resolution}

In this section we present the different resolutions used along the work. We start presenting the lowest resolution, which is England and its nine regions. We then present the one used for the local police forces to then present the one used for the hospital admissions. 

However, we recommend to visit the \href{https://github.com/LeonardoCastro/BritishDrugDynamics/tree/master/tables}{github repository }as some information is replicated in a friendly way. 

\subsection{England and its 9 regions}

These are the most simple and trivial resolutions. England is one of the four nations comprising the United Kingdom and shares borders with Scotland and Wales. England by itself is traditionally divided into 9 regions. These divisions by themselves are the top tier sub-national divisions, and although they do not hold governmental and administrative powers, these regions are often used for statistical and administrative means. These are: 1. East of England, 2. East Midlands, 3. London, 4. North East, 5. North West, 6. South East, 7. South West, 8. West Midlands and 9. Yorkshire and the Humber. 

\subsection{Local Police Forces}

The United Kingdom has a handful of ``British police forces'' as the National Crime Agency, the British Transport Police or the British Borders Police are. Instead, most of the police tasks are taken by local polices acting in a limited area. 

In England there are 39 local polices. Some of them act in unitary local authorities, like Lincolnshire and Northamptonshire Polices, whereas other act in metropolitan regions, like the Metropolitan Police in most of London and the Greater Manchester Police. Also, some polices act in a mixed area comprised of rural and different urban areas. Examples of these are the Thames Valley Police, acting en Oxfordshire, Reading, Milton Keynes, etc., or the Northumbria police acting in Sunderland, Newcastle upon Tyne, Northumbria, etc. 

The list of the different police forces is:
\begin{enumerate}
    \item \textbf{East of England:} Bedfordshire, Cambridgeshire, Essex, Hertfordshire, Norfolk, and Suffolk Polices.
    \item \textbf{East Midlands:} Derbyshire, Leicestershire, Lincolnshire, Northamptonshire, and Nottinghamshire Polices. 
    \item \textbf{London:} Metropolitan Police and the City of London Police.
    \item \textbf{North East:} Cleveland, Durham, and Northumbria Polices.
    \item \textbf{North West:} Cheshire, Cumbria, Great Manchester, Lancashire, and Merseyside Polices.
    \item \textbf{South East:} Hampshire, Kent, Surrey, Sussex, and Thames Valley Polices
    \item \textbf{South West:} Avon and Somerset, Devon and Cornwall, Dorset, Gloucestershire, and Wiltshire Polices.
    \item \textbf{West Midlands:} Staffordshire, Warwickshire, West Mercia, and West Midlands Polices.
    \item \textbf{Yorkshire and the Humber:} Humberside, North Yorkshire, South Yorkshire and West Yorkshire Polices.
\end{enumerate}
Additionally, in the \href{https://github.com/LeonardoCastro/BritishDrugDynamics/tree/master/tables/Police_forces.md}{github repository}, a list of equivalences between the police forces and merged local authorities is shown.

\subsection{Merged local authorities}

In Section \ref{ssec:database} we presented different statistics used throughout this work published from different Governmental offices. For most of them, mainly the ONS, the Home Office and \verb|data.police.gov.uk|, the different resolutions used are consistent during the time interval analysed (2009-2019). However, the hospital admissions data published by NHS digital changed its lower tier territorial divisions in 2012, thus not allowing to have a coherent time series. 

In order to fix this, we created our own lower tier divisions and we call these divisions as \textit{merged local authorities}. This topology of merged local authorities is transferable to the other statistics, allowing us to have a full homogenised database. 

Up until 2012, the NHS was divided in 10 different Strategic Health Areas (similar to the regions described above) and 152 Primary Care Trusts (PCTs) covering England. However, that year the British Parliament passed the Healh and Social Care Act 2012, abolishing SHAs and PCTs, transfering the administrative powers to the 151 Local Authorities in England.

Our homogenisation process involves the detection of the local authorities comprising each PCT, and the detection of PCTs comprising each local authorities. Once done that, the largest number of merged local authorities comprising the 152 PCTs are chosen. The result is a list of 131 merged local authorities.  The equivalence between these, the pre-2012 PCTs and the current local authorities can be found in the \href{https://github.com/LeonardoCastro/BritishDrugDynamics/tree/master/tables/Locations.md}{github repository}.
\end{appendices}

\end{document}